\documentclass[aps,preprint]{revtex4}%
\usepackage{amsfonts}
\usepackage{amsmath}
\usepackage{amssymb}
\usepackage{graphicx}%
\begin{document}
\title{A model for the condensation of a dusty plasma}
\author{P. M. Bellan}
\affiliation{Applied Physics, Caltech, Pasadena CA\ 91125}

\begin{abstract}
A model for the condensation of a dusty plasma is constructed by considering
the spherical shielding layers surrounding a dust grain test particle. The
collisionless region less than a collision mean free path from the test
particle is shown to separate into three concentric layers, each having
distinct physics. The method of matched asymptotic expansions is invoked at
the interfaces between these layers and provides equations which determine the
radii of the interfaces. Despite being much smaller than the Wigner-Seitz
radius, the dust Debye length is found to be physically significant because it
gives the scale length of a precipitous cut-off of the shielded electrostatic
potential at the interface between the second and third layers. Condensation
is predicted to occur when the ratio of this cut-off radius to the
Wigner-Seitz radius exceeds unity and this prediction is shown to be in good
agreement with experiments.

\end{abstract}
\maketitle

\section{Introduction}

Condensation of a dusty plasma \cite{Shukla:2002} into a crystalline state was
proposed by Ikezi in 1986 \cite{Ikezi:1986} and demonstrated experimentally
eight years later by a number of research groups
\cite{Chu:1994,Thomas:1994,Hayashi:1994,Melzer:1994}. The subject has been
reviewed by Morfill et al. \cite{Morfill:2002} and most recently an experiment
to test dusty plasma physics has been set up on the International Space
Station \cite{PKE:2003}.

The original model \cite{Ikezi:1986} for this process was motivated by
Monte-Carlo calculations \cite{Slattery:1980} which predicted that a Coulomb
crystal would form when the Coulomb interaction energy between two adjacent
charged particles in a one-component-plasma exceeded their thermal energy by
some factor. The Coulomb interaction energy for charged particles with density
$n\ $is the electrostatic energy of one particle in the potential of an
adjacent particle located at the Wigner-Seitz interparticle separation
distance \cite{Wigner:1938}%
\begin{equation}
a=\left(  \frac{3}{4\pi n}\right)  ^{1/3}\ . \label{Wigner-Seitz}%
\end{equation}
According to the Monte-Carlo calculations, condensation of charged particles
into a crystal should occur when
\begin{equation}
\ \Gamma=\frac{Z^{2}e^{2}}{4\pi\varepsilon_{0}a\kappa T}\gtrsim
170\ \ \ \label{gamma}%
\end{equation}
where $Z$ is the charge on each particle and $T$ is the temperature of the
particles. As noted by Ikezi \cite{Ikezi:1986}, \ Eq.(\ref{gamma}) could be a
very poor estimate for dusty plasmas (which are a three-component-plasma), but
lacking a better model, Eq.(\ref{gamma}) has often been used as a benchmark
for dusty plasma crystallization experiments. The experiments show
\cite{Melzer:1994} that the actual value of $\Gamma$ required for condensation
is two to three orders of magnitude larger than that predicted by
Eq.(\ref{gamma}). Thus, while Ikezi's original postulate that dusty plasmas
can condense into crystals has been experimentally validated, there has not
been a quantitative model predicting the value of $\Gamma$ necessary for
condensation to occur.

Interactions between adjacent particles in a plasma are intimately related to
the concept of Debye shielding. According to this concept, any plasma particle
can be considered to be a test particle surrounded by a screening cloud of
adjacent particles. The screening completely cancels the test particle field
at distances much greater than the Debye length. \ Screening may be
accomplished by adjacent particles of either the same polarity as the test
charge or opposite polarity, but is subject to the constraint that the test
particle cannot be moving faster than the thermal velocity of the shielding
particles \cite{Nicholson}. For example, if the test particle is an electron,
it is shielded by the repulsion of other electrons in the presence of a
uniform neutralizing ion background, but it is not shielded by ions because it
is moving too fast for ions to respond. On the other hand, ions are shielded
by both electrons and ions. In a dusty plasma one might thus reasonably expect
dust grains to be shielded by electrons, ions, and other dust grains.

The standard model of Debye shielding is based on the Boltzmann relation, an
equilibrium solution to the fluid equation of motion for each species $\sigma$
such that the force due to the electric field balances the force due to the
gradient of an isotropic scalar pressure, i.e.,%
\begin{equation}
0=-n_{\sigma}q_{\sigma}\nabla\phi-\nabla P_{\sigma}. \label{Boltz}%
\end{equation}
Three critical assumptions are intrinsic to the standard model of Debye
shielding, namely: (i)\ it is assumed that the plasma is sufficiently
collisional that the concept of an isotropic scalar pressure $P_{\sigma}=$
$n_{\sigma}\kappa T_{\sigma}$ is valid, (ii)\ it is assumed that a Boltzmann
dependence $n_{\sigma}=n_{\sigma0}\exp(-q_{\sigma}\phi/\kappa T_{\sigma})$
exists relating the local density $n_{\sigma}$ to the system-averaged density
$n_{\sigma0}$, and (iii) it is assumed that $\left\vert q_{\sigma}\phi/\kappa
T_{\sigma}\right\vert <<1\,$so that the Boltzmann relationship may be
linearized giving $n_{\sigma}/n_{\sigma0}=\ 1-q_{\sigma}\phi/\kappa T_{\sigma
}$. The standard model for Debye shielding of a test particle with charge
$q_{T}$ results when the linearized Boltzmann relationships of the various
species are substituted into Poisson's equation giving the Yukawa-type
solution $\phi(r)=q_{T}\exp(-r/\lambda_{D})/4\pi\varepsilon_{0}r$ where
\begin{equation}
\frac{1}{\lambda_{D}^{2}}=\sum\frac{1}{\lambda_{D\sigma}^{2}}
\label{Debye sum}%
\end{equation}
and
\begin{equation}
\lambda_{D\sigma}^{2}=\frac{\varepsilon_{0}\kappa T_{\sigma}}{n_{\sigma
0}q_{\sigma}^{2}}. \label{Debye}%
\end{equation}
The summation in Eq.(\ref{Debye sum}) is restricted to species that
participate in the shielding and so excludes all species having thermal
velocity slower than the species of the test particle.

When $r\rightarrow0,\,\ $the Yukawa solution diverges, violating assumption
(iii)$\ $ that $\left\vert q_{\sigma}\phi/\kappa T_{\sigma}\right\vert <<1$
and causing the standard model to fail to be internally self-consistent. This
failure of the standard model of Debye shielding has been noted previously,
e.g., see Lampe, Joyce and Ganguli \cite{Lampe:2000} for criticism regarding
assumptions (i)-(iii). In addition, Hansen and Fajans \cite{Hansen:1995} have
shown that trapping can affect Debye shielding in a pure electron plasma while
Goree \cite{Goree:1992}, Zobnin et al. \cite{Zobnin:2000}, and Lampe et al.
\cite{Lampe:2003} have shown that trapping of ions can affect shielding of a
dust grain.

The issue of how to treat Debye shielding when $\left\vert q_{\sigma}%
\phi/\kappa T_{\sigma}\right\vert >>1$ is especially critical for the dust
condensation problem, because $\left\vert q_{\sigma}\phi/\kappa T_{\sigma
}\right\vert $ is essentially the same as $\Gamma$. Consensus does not exist
on how to address this issue. \ 

Furthermore, the form of Eq.(\ref{Debye sum}) is such that the sum on the
right hand side is dominated by the term having the smallest $\lambda
_{D\sigma}^{2}$ and, since dust particles are both cold and highly charged,
the dust Debye length is typically much smaller than both the electron and ion
Debye lengths. One might expect that the system Debye length $\lambda_{D}$
should be very nearly the dust Debye length $\lambda_{Dd}$, but this point of
view has usually been rejected. The dust Debye length $\lambda_{Dd}$ is
typically so small that it is less than $a$, and questions have been raised as
to whether such a short shielding length has physical significance since the
standard Debye argument is based on the implicit assumption that there is a
statistically large number of particles in a sphere having the Debye radius.
This is clearly not true for dust particles in a sphere with radius
$\lambda_{Dd}$ if $\lambda_{Dd}$ is less than $a.$ Nevertheless, Wang and
Bhattarcharjee \cite{Wang:1995} and also Otani and Bhattarcharjee
\cite{Otani:1997} argued that some sort of shielding does occur at the scale
of $\lambda_{Dd}$ but the only support for this point of view was
demonstration \cite{Otani:1997} of some time-averaged correlation effects at
the scale of $\lambda_{Dd}$ in a one-dimensional numerical simulation that
would only crystallize if artificially annealed. Most other authors ignore
dust self-shielding on the presumption that the Debye shielding concept does
not make sense when a Debye length is smaller than $a.$

We present here a model for a dusty plasma on the verge of condensation. This
model takes into account both collisional and collisionless behavior in three
dimensional geometry, avoids inappropriate use of fluid theory, shows that the
dust Debye length has important physical significance even though it is much
smaller than $a,$ and predicts a condensation threshold in good agreement with
experimental measurements. The derivation identifies four physically distinct
concentric regions surrounding a test charge. The ions are collisionless in
the innermost three regions but collisional in the outer, fourth region (the
collisionless nature of ions in the inner three regions is consistent with the
assumptions inherent in dust grain charging theory). The method of matched
asymptotic expansions is used to locate the two interfaces between the first
three regions and knowledge of these interface locations is then used to give
the criterion for condensation. \ The $T_{e}>>T_{i},T_{d}$ temperature regime
of typical dusty plasma condensation experiments is assumed $\ $ and ions are
assumed to be singly charged (the theory could be extended to arbitrary
temperatures without great difficulty, but this would unnecessarily complicate
the model).

The paper is organized as follows:\ Section II reviews relevant aspects of
dust charging theory and sets up a dimensionless parameter space suitable for
comparing the model to experiments. Section III uses collisionless Vlasov
theory to calculate the ion, electron, and dust grain densities and shows that
when $\left\vert e\phi/\kappa T_{i}\right\vert >1$, the ion density differs
from the Boltzmann model; this difference demonstrates the inappropriateness
of fluid models in this regime and resolves the paradox associated with
divergence of the Yukawa solution at small $r.$ Section IV\ shows that the
vicinity of a test particle can be divided into three concentric spherical
regions each having distinct physics determined by the magnitude of
$\left\vert e\phi/\kappa T_{i}\right\vert $. Section V\ derives approximate
solutions to the Vlasov-Poisson system for these three regions and Section VI
derives matching conditions across the two interfaces between the three
regions. Section VII uses the matching conditions to deduce a condition for
dust condensation and compares the model predictions with experiments. Section
VIII provides a summary and discussion.

\section{Dust Charging and Dusty Plasma Parameter Space}

Two independent parameters characterize the dust grains in a dusty
plasma:\ the grain radius $r_{d}$ and the Wigner-Seitz radius $a.$ In order to
develop a model based on dimensionless parameters, the ion Debye length
\begin{equation}
\lambda_{Di}=\sqrt{\frac{\varepsilon_{0}\kappa T_{i}}{n_{i0}e^{2}}}
\label{LambdaDi}%
\end{equation}
will be used as the `yardstick' by which all lengths are measured. A bar will
be used to denote lengths normalized to the ion Debye length so that the
normalized Wigner-Seitz radius, for example, is%
\begin{equation}
\bar{a}^{\ }=\ \frac{1}{\lambda_{Di}}\left(  \frac{3}{4\pi n_{d0}}\right)
^{1/3}\ \ \ . \label{a-norm}%
\end{equation}
The two quantities $\bar{a}$ and $\bar{r}_{d}$ constitute the coordinates for
a dimensionless dusty plasma parameter space.

In order to\ avoid confusing minus signs, the electrostatic potential $\phi$
will be replaced by the positive dimensionless variable
\begin{equation}
\psi=-\frac{e\phi}{\kappa Ti} \label{psi defn}%
\end{equation}
and $\psi_{d}$ will denote the potential on the surface of a dust grain. Thus,
positive $\psi$ attracts ions but repels both electrons and dust grains.

When dust grains are placed in an electron-ion plasma, some fraction of the
electrons attach to the dust grain surface, causing the dust grains to become
negatively charged and reducing the density of free electrons. The
quantitative theory of dust charging, summarized in Ref. \cite{Shukla:2002},
combines collisionless Vlasov theory with an analysis of trajectories of
individual particles as they approach a finite radius charged sphere. The
particle trajectories are assumed to be governed by orbital-motion-limited
(OML) theory \cite{Mott:1926,Allen:1956,Shukla:2002OML} wherein particle
trajectories are assumed to be collisionless and completely determined by
considerations of conservation of angular momentum and conservation of energy.
There has been some question \cite{Melzer:1994} about the extent to which the
standard dust charging model applies to dust grains in an electrode sheath,
the typical situation for terrestrial dusty plasma condensation experiments,
but not for the zero-gravity dusty plasma condensation experiment on the
International Space Station. We assume in this paper that the standard dust
charging model is applicable so that the effect, if any, of electrode sheaths
on dust charging is small.

The standard dust charging model shows that dust grain charging is governed by
the dimensionless parameter%
\begin{equation}
P=4\pi n_{d0}\lambda_{Di}^{2}r_{d}=4\pi n_{d0}\lambda_{Di}^{3}\bar{r}%
_{d}=\frac{3\bar{r}_{d}}{\bar{a}^{3}}\ \label{charging param}%
\end{equation}
where $P$ has the functional dependence%
\begin{equation}
P=\frac{1}{\psi_{d}}-\left(  1+\frac{1}{\psi_{d}}\right)  \sqrt{\frac
{m_{e}T_{i}}{m_{i}T_{e}}}\exp(\psi_{d}T_{i}/T_{e}). \label{P dep}%
\end{equation}

Global quasineutrality gives
\begin{equation}
Z_{d}n_{d0}+n_{e0}=n_{i0} \label{qn}%
\end{equation}
where $Z_{d}$ is the number of electrons captured by a dust grain. We define
the electron capture factor%
\begin{equation}
\alpha=Z_{d}n_{d0}/n_{i0} \label{define alpha}%
\end{equation}
so that $\alpha=1$ corresponds to having all the electrons attached to the
dust grains while $\alpha=0$ corresponds to having no electrons attached to
the dust grains. The quasineutrality condition, Eq. (\ref{qn}), can thus be
expressed as%
\begin{equation}
\alpha+\frac{n_{e0}}{n_{i0}}=1 \label{qn alpha}%
\end{equation}
and dust charging theory \cite{Shukla:2002a} shows that
\begin{equation}
\alpha=P\psi_{d}. \label{alpha dep}%
\end{equation}
Since $\psi_{d}$ and $\alpha$ are functions of $P,$ they have functional
dependence $\alpha=\alpha(\bar{a},\bar{r}_{d})$ and $\psi_{d}=\psi_{d}(\bar
{a},\bar{r}_{d}).$

Combining Eqs. (\ref{charging param}), (\ref{define alpha}) and
(\ref{alpha dep}) shows that%
\begin{equation}
\frac{Z_{d}}{4\pi n_{i0}\lambda_{Di}^{3}}=\ \bar{r}_{d}\psi_{d}\ \label{Zd}%
\end{equation}
so that $Z_{d}$ becomes large if $\psi_{d}$ is finite, $\bar{r}_{d}$ is not
infinitesimal, and $4\pi n_{i0}\lambda_{Di}^{3}$ is large. Equation (\ref{Zd})
is just the normalized version of the potential $\phi_{d}=-Z_{d}%
e/4\pi\varepsilon_{0}r_{d}$ of a sphere of radius $r_{d}$ with surface charge
$-Z_{d}e$. This result is actually slightly incorrect for a shielded dust
grain, because, as shown in the next paragraph, the shielding cloud
surrounding a dust grain depresses the potential at the grain surface to a
value below the value given by Eq.(\ref{Zd}).

To understand this potential depression effect due to shielding, consider the
potential $\phi$ on the surface of a sphere with charge $Q$ and radius
$r_{sphere}$ surrounded by a shell of shielding charge $-Q$ at radius
$r_{shell}.$ The potential on the surface of the shielded sphere is given by
\begin{equation}
\phi(r_{sphere})=\int_{\infty}^{r_{sphere}}dr\frac{\partial\phi}{\partial
r}=\frac{Q}{4\pi\varepsilon_{0}}\left(  \frac{1}{r_{sphere}}-\frac
{1}{r_{shell}}\right)  , \label{shielded pot}%
\end{equation}
a result obtained by taking into account the contributions to $\partial
\phi/\partial r$ from both the sphere and its shielding charge. The ratio of
the surface potential of the shielded sphere to the potential of an identical
unshielded sphere is $\phi(r_{sphere})/\phi_{vac}=1-r_{sphere}/r_{shell}$
where $\phi_{vac}$ is the surface potential of the unshielded sphere. If
$r_{shell}-r_{sphere}\simeq\lambda_{D}$ where $\lambda_{D}$ is the nominal
Debye length, then $\phi(r_{sphere})/\phi_{vac}\simeq\lambda_{D}/\left(
r_{sphere}+\lambda_{D}\right)  $ so the potential of the shielded sphere will
be greatly depressed from its vacuum value if $r_{sphere}>>\lambda_{D}.$ This
indicates that being highly charged is insufficient for a dust grain to have a
large potential; \ it also needs to have $\bar{r}_{d}<<1.$ The model of dust
charging thus has the implicit assumption that $\bar{r}_{d}$ is small compared
to unity and this assumption will be made in the remainder of this paper.

Figures 1(a)-(c)\ show contours of constant $\ \psi_{d},$ $\alpha,$ and
$\ Z_{d}/4\pi n_{i0}\lambda_{Di}^{3}$ as determined by
Eqs.(\ref{charging param}), (\ref{P dep}), (\ref{alpha dep}) and (\ref{Zd})
for the parameters of Ref. \cite{Chu:1994}, a typical dust crystallization
experiment. Since dust grains in a given experiment have a fixed ratio
$\ \bar{r}_{d}/\bar{a},$ a specific experiment is characterized by a sloping
straight line in $\bar{a},\bar{r}_{d}$ parameter space. Moving up and to the
right along such a line corresponds to making $\lambda_{di}$ smaller whereas
moving down and to the left corresponds to making $\lambda_{di}$ larger.
Densities in an experiment are typically measured by \ Langmuir probes which
have an uncertainty of -50\%, +100\%, so that the density of $n_{i}=10^{9}$
cm$^{-3}$ reported in \cite{Chu:1994} would actually be in the range
$5\times10^{8}$ cm$^{-3}<n_{i}<2\times10^{9}$ cm$^{-3}.$ This factor of four
range of densities \cite{Thomas:1994} corresponds to the straight line segment
labeled `expt' in Figs.1(a)-(c). This line has a slope given by $\bar{r}%
_{d}/\bar{a}=r_{d}/a.$ The left end of this line is the point in parameter
space calculated using the lower estimate for the density, while the right end
corresponds to using the upper estimate for the density. The length of this
line effectively represents the density measurement error bar. To the extent
that charging theory is correct, the range of possible values of $\alpha,$
$\psi_{d},$ and $Z_{d}/4\pi n_{i0}\lambda_{Di}^{3}$ for the experiment are
given by the intersection of the contours with this `expt' line.%

%TCIMACRO{\FRAME{fhFU}{6.6666in}{4.5878in}{0pt}{\Qcb{ Dusty plasma parameter
%space for Chu and I experiment (density range indicated by short line labeled
%expt):\ horizontal axis is $\bar{a}=a/\lambda_{di},$ vertical axis is $\bar
%{r}_{d}=r_{d}/\lambda_{di}$; (a)\ shows contours of constant $\psi_{d}$, (b)
%shows $\ $contours of constant $\alpha$, (c) shows contours of constant
%$Z_{d}/4\pi n_{i0}\lambda_{di}^{3}$.}}{}{049406php1.eps}%
%{\special{ language "Scientific Word";  type "GRAPHIC";
%maintain-aspect-ratio TRUE;  display "USEDEF";  valid_file "F";
%width 6.6666in;  height 4.5878in;  depth 0pt;  original-width 6.7057in;
%original-height 4.6052in;  cropleft "0";  croptop "1";  cropright "1";
%cropbottom "0";  filename '049406PHP1.eps';file-properties "XNPEU";}}}%
%BeginExpansion
\begin{figure}
[h]
\begin{center}
\includegraphics[
height=4.5878in,
width=6.6666in
]%
{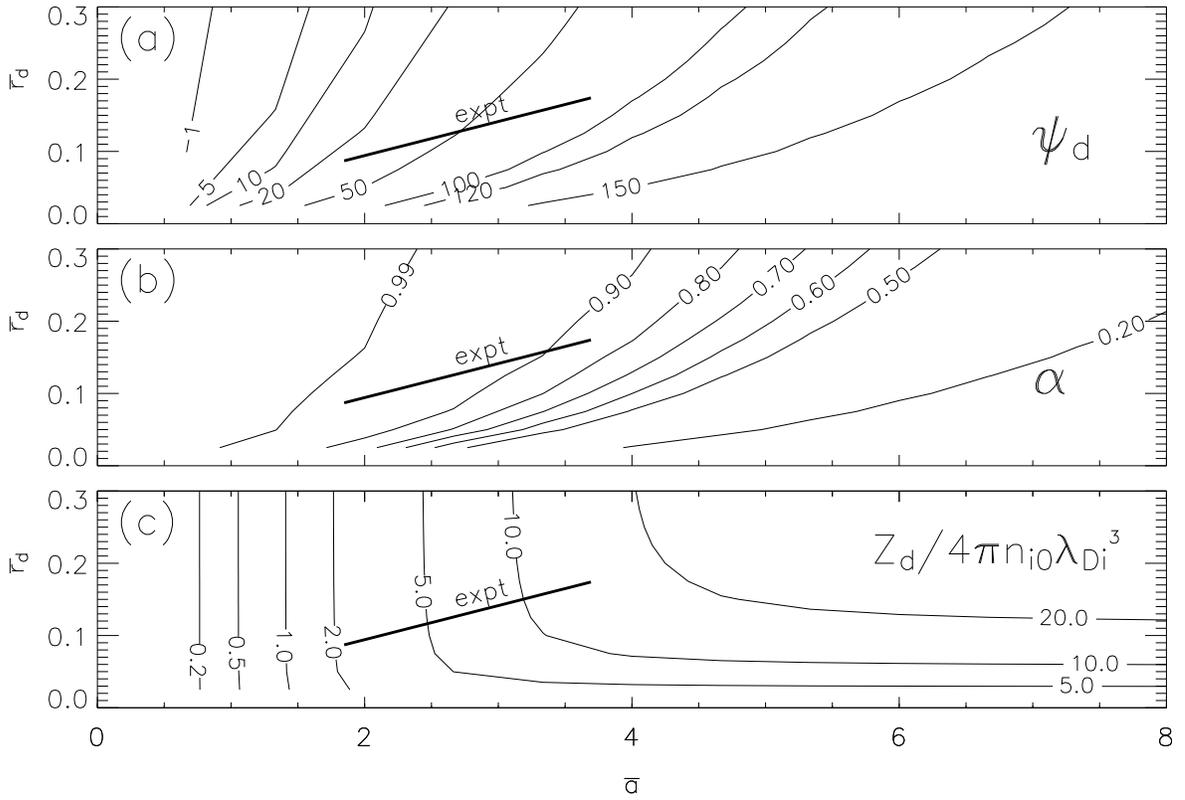}%
\caption{ Dusty plasma parameter space for Chu and I experiment (density range
indicated by short line labeled expt):\ horizontal axis is $\bar{a}%
=a/\lambda_{di},$ vertical axis is $\bar{r}_{d}=r_{d}/\lambda_{di}$;
(a)\ shows contours of constant $\psi_{d}$, (b) shows $\ $contours of constant
$\alpha$, (c) shows contours of constant $Z_{d}/4\pi n_{i0}\lambda_{di}^{3}$.}%
\end{center}
\end{figure}
%EndExpansion

\section{Vlasov Model of Charged Particle Density in the Presence of a
Potential}

A typical dust grain will be considered as a test particle inserted in a
plasma consisting of electrons, ions, and other dust grains. The origin of a
spherical coordinate system will be defined to be at the center of this test
particle. \ Typical dust condensation experiments have neutral pressures
$\sim10^{2}$ Pa (corresponding to a neutral density $n_{n}$ $\allowbreak
\simeq3\ \times10^{16}$ $\ $cm$^{-3}$). \ Since neutral cross-sections are
$\sigma\sim3\times10^{-16}$ cm$^{2}$, the mean free path for ion-neutral
collisions is $l_{mfp}=(n_{n}\sigma)^{-1}\sim1$ mm$\allowbreak\allowbreak\ $
which is at least an order of magnitude larger than shielding scale lengths.
The last collision experienced by an ion in the vicinity of the test particle
will have occurred outside a sphere having a normalized diameter of the order
of $\bar{l}_{mfp}$; such a sphere is shown schematically in Fig. 2 and lies at
the interface between regions 3 and 4. Thus ions can be considered
collisionless inside regions 1 to 3 and collisional in region 4; the details
of regions 1 to 3 will be discussed later. This separation of space into
collisional and collisionless regions is similar to the arguments used in dust
grain charging theory (the OML assumption underlying dust grain charging
theory is based on angular momentum conservation which can only be true if a
particle has no collisions).%

%TCIMACRO{\FRAME{fhFU}{2.5986in}{3.4695in}{0pt}{\Qcb{Sketch of concentric
%regions surrounding a dust grain test particle of normalized radius $\bar
%{r}_{d}$. Diameter of outermost dashed circle is $\bar{l}_{mfp}$, the
%normalized mean free path for collisions, so plasma is collisional in region 4
%outside this circle. Regions 1, 2, and 3 are collisionless and have interfaces
%on the dashed circles having normalized radii $\bar{r}_{i}$ and $\bar{r}_{o}%
%.$}}{}{049406php2.eps}{\special{ language "Scientific Word";  type "GRAPHIC";
%maintain-aspect-ratio TRUE;  display "USEDEF";  valid_file "F";
%width 2.5986in;  height 3.4695in;  depth 0pt;  original-width 2.5703in;
%original-height 3.4404in;  cropleft "0";  croptop "1";  cropright "1";
%cropbottom "0";  filename '049406PHP2.eps';file-properties "XNPEU";}}}%
%BeginExpansion
\begin{figure}
[h]
\begin{center}
\includegraphics[
height=3.4695in,
width=2.5986in
]%
{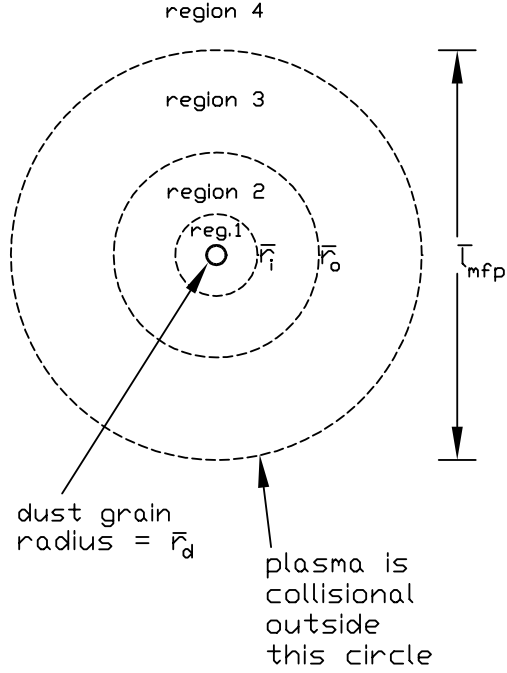}%
\caption{Sketch of concentric regions surrounding a dust grain test particle
of normalized radius $\bar{r}_{d}$. Diameter of outermost dashed circle is
$\bar{l}_{mfp}$, the normalized mean free path for collisions, so plasma is
collisional in region 4 outside this circle. Regions 1, 2, and 3 are
collisionless and have interfaces on the dashed circles having normalized
radii $\bar{r}_{i}$ and $\bar{r}_{o}.$}%
\end{center}
\end{figure}
%EndExpansion

Electrostatic potential is undefined with respect to a constant; following
convention, we choose this constant such that $\phi=0$ at infinity. Collisions
$\ $make the distribution function Maxwellian in region 4 and this provides a
boundary condition for the collisionless distribution function in regions 1 to
3. The distribution function in regions 1 to 3 must satisfy the collisionless
Vlasov equation and so must be must be a function of constants of the motion
\cite{Watson:1956,Nicholson:motionconstants}. The relevant constant of the
motion here is the particle energy $W=m_{\sigma}v^{2}/2+q_{\sigma}\phi$, and
so the distribution function in regions 1 to 3 is
\begin{equation}
f_{\sigma}(\mathbf{r,v)}=n_{\sigma0}\left(  \frac{m_{\sigma}}{2\pi\kappa
T_{\sigma}}\right)  ^{3/2}\exp\left(  -\frac{m_{\sigma}v^{2}/2+q_{\sigma}%
\phi(\mathbf{r)}}{\kappa T_{\sigma}}\right)  \ . \label{f soln}%
\end{equation}
This is the right choice because $f_{\sigma}(\mathbf{r,v)}$ is not only a
function of a constant of the motion $W$ but also joins smoothly to the region
4\ Maxwellian solution where $\phi=0$ and the plasma is collisional.

Electrons and dust grains experience a repulsive force upon approaching the
negatively charged dust grain test particle and so are slowed down with some
particles being slowed down to zero velocity and reflecting. Thus, electrons
or dust grains near the dust grain test particle can have zero velocity. The
respective electron and dust grain densities in the vicinity of the dust grain
test particle are thus given by
\begin{align}
n_{e}  &  =\int_{0}^{\infty}4\pi v^{2}dvf_{e}(\mathbf{r,v)}=n_{e0}\exp\left(
\frac{\ e\phi(\mathbf{r)}}{\kappa T_{e}}\right)  ,\label{e density}\\
\ \text{ }n_{d}  &  =\int_{0}^{\infty}4\pi v^{2}dvf_{d}(\mathbf{r,v)}%
=n_{d0}\exp\left(  \frac{\ Z_{d}e\phi(\mathbf{r)}}{\kappa T_{d}}\right)  .
\label{d density}%
\end{align}
Using Eq.(\ref{psi defn}), the electron and dust grain densities normalized to
their average values are%
\begin{equation}
\frac{n_{e}}{n_{e0}}=\ \exp\left(  -\psi T_{i}/T_{e}\right)  \label{ne}%
\end{equation}
and%
\begin{equation}
\frac{n_{d}}{n_{d0}}=\ \exp\left(  -Z_{d}\psi T_{i}/T_{d}\right)  . \label{nd}%
\end{equation}
These densities are identical to the Boltzmann relation and so demonstrate
that collisionless kinetic theory agrees with fluid theory for negatively
charged particles near a negatively charged test particle.

Ion behavior is fundamentally different because ions, being positive, are
accelerated as they approach the negatively charged dust grain test particle;
this means that there are no zero velocity ions near a dust grain. The slowest
ion is one that has fallen into the negative $\phi$ well with zero initial
velocity at the edge of the well and the velocity of such an ion will satisfy
\begin{equation}
m_{i}v^{2}/2+e\phi=0. \label{ion energy}%
\end{equation}
Using Eq.(\ref{psi defn}) it is seen that this minimum possible ion velocity
can be expressed as
\begin{equation}
\ v_{\min}\ =\sqrt{\frac{2\kappa T_{i}}{m_{i}}\psi}. \label{vmin}%
\end{equation}

Evaluation of the ion density in the vicinity of the dust grain test particle
therefore requires invoking a lower limit at $v_{\min}$ for the velocity
integration over the distribution function. The resulting ion density is thus%
\begin{align}
n_{i}  &  =\int_{v_{\min}}^{\infty}4\pi v^{2}dvf_{i}(\mathbf{r,v)}\nonumber\\
&  =n_{i0}\left(  \frac{m_{i}}{2\pi\kappa T_{i}}\right)  ^{3/2}\exp\left(
\psi\right)  \int_{\sqrt{2\kappa T_{i}\psi/m_{i}}}^{\infty}4\pi v^{2}%
dv\exp\left(  -\frac{m_{i}v^{2}}{2\kappa T_{i}}\right)  . \label{ni calc}%
\end{align}
By defining $\xi=v/\sqrt{2\kappa T_{i}/m_{i}},$ the normalized ion density can
be expressed as
\begin{align}
\frac{n_{i}}{n_{i0}}  &  =\ \ \ \frac{4e^{\psi}}{\sqrt{\pi}}\int_{\sqrt{\psi}%
}^{\infty}\xi^{2}d\xi\exp\left(  -\xi^{2}\right) \nonumber\\
&  =\ e^{\psi}\left(  1-\operatorname{erf}\sqrt{\psi}\right)  +\frac{2}%
{\sqrt{\pi}}\sqrt{\psi} \label{ni}%
\end{align}
where
\begin{equation}
\operatorname{erf}z=\frac{2}{\sqrt{\pi}}\int_{0}^{z}e^{-\xi2}d\xi\label{erf}%
\end{equation}
is the error function. The second line in Eq.(\ref{ni}) is obtained using the
identity
\begin{equation}
\int_{\sqrt{\psi}}^{\infty}d\xi\frac{d}{d\xi}\left(  \xi\exp\left(  -\xi
^{2}\right)  \right)  =\int_{\sqrt{\psi}}^{\infty}d\xi\exp(-\xi^{2}%
)-\ 2\int_{\sqrt{\psi}}^{\infty}d\xi\ \xi^{2}\exp(-\xi^{2}). \label{identity}%
\end{equation}

\bigskip For small arguments, the error function may be approximated%
\begin{equation}
\lim_{z\rightarrow0}\operatorname{erf}z=\frac{2}{\sqrt{\pi}}\left(
z-\frac{z^{3}}{3}\right)  . \label{erf limit}%
\end{equation}
Thus for $\psi<<1\,$\ and hence $\sqrt{\psi}<<1,\ $the normalized ion density
has the form%
\begin{align}
\frac{n_{i}}{n_{i0}}  &  =\ \ \ \left(  1+\psi+\frac{1}{2}\psi^{2}\right)
\left(  1-\frac{2}{\sqrt{\pi}}\left(  \sqrt{\psi}-\frac{\psi^{3/2}}{3}\right)
\right)  +\frac{2}{\sqrt{\pi}}\sqrt{\psi}\nonumber\\
&  \simeq1+\psi\label{ion small psi}%
\end{align}
which is the same as the Boltzmann result given by fluid theory.

However, because%
\[
\lim_{\psi\rightarrow\infty}e^{\psi}\left[  1-\operatorname{erf}\left(
\sqrt{\psi}\right)  \right]  =0,
\]
the normalized ion density when $\psi>>1$ is
\begin{equation}
\frac{n_{i}}{n_{i0}}\simeq\frac{2}{\sqrt{\pi}}\sqrt{\psi}; \label{inner ion}%
\end{equation}
this is much smaller than the fluid theory Boltzmann relation prediction that
$n_{i}/n_{i0}=\exp(\psi)$. Equation (\ref{inner ion}) thus demonstrates a
failure of fluid theory and its associated Boltzmann relationship when
$\psi>>1$. This failure occurs because the concept of ion pressure no longer
makes sense when $\psi>>1$. The pressure concept is based on the assumption
that particles have an isotropic Gaussian distribution of random velocities
about some mean velocity whereas when $\psi>>1,$ ions in reality are falling
into a deep potential well and do not have a random distribution of velocities
about some mean velocity. Equation (\ref{ni}) and the distinction between its
small and large $\psi$ limits have been previously discussed by Laframboise
and Parker \cite{Laframboise:1973} in the context of electrostatic probes and
by Lampe, Joyce and Ganguli \cite{Lampe:2000} in the context of dusty plasmas.

The lower limit of the integral in Eq.(\ref{ni calc}) causes the ion
distribution to have an $r-v$ phase-space `hole' in the vicinity of the dust
grain since $f(r,v)=0$ for velocities below $v_{\min}$. It has been argued by
Bernstein and Rabinowitz \cite{Bernstein:1959}, Laframboise and Parker
\cite{Laframboise:1973}, and Lampe \cite{Lampe:2001} that for a certain class
of radial potential profiles, another sort of phase-space hole can also exist.
This additional hole results from a rather subtle barrier that can occur
because for a certain range of the angular momentum $J$ the effective
potential $U_{eff}(r)=q\phi(r)+J^{2}/2mr^{2}$ can have a small local maximum.
This barrier prevents access to small $r$ by particles having a certain range
of $W$ and $J$. If such a barrier exists, the radial ion density profile will
differ somewhat from the predictions of OML\ theory, because the ions that
cannot pass by this barrier will have a radial turning point at a larger
radius than predicted by OML. However, since Poisson's equation shows that
$\psi$ is essentially a double integral of the net charge density up to a
radius $r,$ changes in the turning point of small classes of ions should not
have a major effect on the $\psi$ profile, i.e., small corrections to the
OML\ model should not result in a significant collective effect. Lampe
\cite{Lampe:2001} has shown that the error introduced by omission of
consideration of these centrifugal force barriers is very small for dusty
plasmas and so we will ignore this correction to OML\ theory.

Another correction to OML theory results from consideration of ion capture by
the dust grain which also causes a hole in phase-space
\cite{Allen:1956,Bernstein:1959}. As shown in Ref.\cite{Bernstein:1959}
capture of ions by the dust grain reduces the number of ions moving radially
outward from the dust grain in comparison to the limiting situation where the
dust grain does not capture any ions so that all ions are perfectly reflected
from the dust grain. Taking into account the reduction in the number of
outward moving ions compared to inward moving ions would require replacing the
distribution function prescribed by Eq.(\ref{f soln}) by a distribution
function of the form \cite{Bernstein:1959} $f=f_{+}+f_{-}$ where \ $f_{+}(W)$
is the phase space density of ions moving radially outwards from the dust
grain and $f_{-}(W)$ is the phase space density of ions moving radially
inwards. If ions are perfectly reflected at the dust grain then $f_{+}%
=f_{-}\ $\ in which case Eq.(\ref{f soln}) is appropriate, but if ions are
captured by the dust grain then $f_{+}$ \thinspace$<$ $f_{-}$ and a more
complicated prescription than Eq.(\ref{f soln}) would have to be used. We will
assume that the fraction of ions incident at $\bar{r}_{o}$ which are captured
by the dust grain is so small that Eq.(\ref{f soln}) is a reasonably accurate
prescription for the ion phase space density. We are thus assuming that the
$r$ projection of ion motion has a reflecting trajectory so that there are
equal numbers of ions moving radially inwards and outwards in the dust grain
shielding cloud. This assumption will be validated later.

Finally, we will also ignore ion trapping
\cite{Goree:1992,Zobnin:2000,Lampe:2003}, but will later make some brief
comments about the extent to which trapping might be important.

\section{Identification of three regions for the potential}

Poisson's equation
\begin{equation}
\nabla^{2}\phi=-\frac{e}{\varepsilon_{0}}\left(  n_{i}-n_{e}-Z_{d}%
n_{d}\right)  \ \label{Poisson}%
\end{equation}
relates the densities of the various species to the electrostatic potential.
Assuming spherical symmetry about the dust grain test particle and using
Eq.(\ref{psi defn}), Poisson's equation can be recast as%
\begin{equation}
\frac{1}{\bar{r}^{2}}\frac{\partial}{\partial\bar{r}}\left(  \bar{r}^{2}%
\frac{\partial\psi}{\partial\bar{r}}\right)  =\frac{n_{i}}{n_{i0}}%
-\frac{n_{e0}}{n_{i0}}\frac{n_{e}}{n_{e0}}-Z_{d}\frac{n_{d0}}{n_{i0}}%
\frac{n_{d}}{n_{d0}}. \label{Poisson norm}%
\end{equation}
Using Eqs.(\ref{ne}), (\ref{nd}), (\ref{ni}) for the normalized densities and
also Eq.(\ref{define alpha}), the normalized Vlasov/Poisson system becomes%
\begin{equation}
\frac{1}{\bar{r}^{2}}\frac{\partial}{\partial\bar{r}}\left(  \bar{r}^{2}%
\frac{\partial\psi}{\partial\bar{r}}\right)  =\ \underset{\text{ions}%
}{\underbrace{e^{\psi}\left(  1-\operatorname{erf}\left(  \sqrt{\psi}\right)
\right)  +\frac{2}{\sqrt{\pi}}\sqrt{\psi}}}-\underset{\text{electrons}%
}{\underbrace{\left(  1-\alpha\right)  \exp\left(  -\frac{\psi T_{i}}{T_{e}%
}\right)  }}-\underset{\text{dust}}{\underbrace{\alpha\exp\left(  -\bar{Z}%
\psi\right)  }} \label{grand PE}%
\end{equation}
where $\bar{Z}=Z_{d}T_{i}/T_{d}$ is presumed to be large compared to unity
since the dust grain is highly charged and $T_{i}>T_{d}.$ Equation
(\ref{grand PE}) is a nonlinear ordinary differential equation for $\psi$ and
is consistent with the collisionless Vlasov equation. Since the densities were
obtained using the collisionless Vlasov equation, this system will be called
the Vlasov/Poisson system to distinguish it from the fluid/Poisson system.

We now argue that three distinct regions exist for $\psi$ such that in each
region the Vlasov/Poisson system has a different form. The location of these
regions is sketched in Fig.2 and, going outwards from the surface of the dust
grain test particle, these regions and their interfaces are defined by:

Region 1 is where $\ \psi_{d}>$ $\psi>1$ and exists because the grain
potential $\psi_{d}$ is large compared to unity [see Fig.1(a)]. Region 1 is a
sheath-like inner region where the ion density has the non-Boltzmann behavior
given by Eq.(\ref{inner ion}).

Region 2 is where $1>\psi>1/Z_{d}$ and is depleted of dust grains. Shielding
in this region is provided mainly by ions.

Region 3 is where $1/Z_{d}>\psi$ and this region extends to region 4 where
collisions set in and where the potential goes to zero. Shielding in region 3
is done mainly by dust grains and this shielding takes place over a very short
characteristic length, causing an extremely sharp cut-off of the potential.

The radii of the respective interfaces between regions 1 and 2 and between
regions 2 and 3 will be called $\bar{r}_{i}$ and $\bar{r}_{o}$ as indicated in
Fig.2 (the subscripts $i~$and $o$ stand for inner and outer interfaces). The
values of $\bar{r}_{i}$ and $\bar{r}_{o}$ will be unknowns to be solved for;
determining these radii is the crux of the problem.

\section{Approximate solutions to the Vlasov/Poisson system for the three
collisionless regions}

The three collisionless regions will now be discussed going from the outermost
(region 3) to the innermost (region 1).

\subsection{Region 3 solution: $\ \psi<1/\bar{Z}$}

In region 3, Eq.(\ref{grand PE}) can be approximated as%
\begin{align}
\frac{1}{\bar{r}^{2}}\frac{\partial}{\partial\bar{r}}\left(  \bar{r}^{2}%
\frac{\partial\psi}{\partial\bar{r}}\right)   &  =\ \underset{\text{ions}%
}{\underbrace{1+\psi}}-\underset{\text{electrons}}{\underbrace{\left(
1-\alpha\right)  \left(  1-\ \frac{\psi T_{i}}{T_{e}}\right)  }}%
-\underset{\text{dust}}{\underbrace{\alpha\left(  1-\bar{Z}\psi\right)  }%
}\nonumber\\
&  =\ \ \ \left(  1+\left(  1-\alpha\right)  \frac{T_{i}}{T_{e}}+\alpha\bar
{Z}\right)  \psi. \label{outer PE}%
\end{align}
Using $\ $ $T_{e}>>T_{i},$ this has the Yukawa-type solution
\begin{equation}
\psi_{3}=\frac{\bar{r}_{o}}{\bar{Z}\bar{r}}\exp\left(  -\sqrt{\alpha\bar
{Z}\ +1\ }\left(  \bar{r}-\bar{r}_{o}\right)  \right)  \ .
\label{small psi soln}%
\end{equation}

The coefficient in Eq.(\ref{small psi soln}) has been chosen so that $\psi
_{3}=1/\bar{Z}$ at $\bar{r}=\bar{r}_{o}.$ The effective shielding length in
region 3 is the dust Debye length%
\begin{equation}
\lambda_{Dd}=\lambda_{Di}/\sqrt{1+\alpha\bar{Z}} \label{dust Debye}%
\end{equation}
which is much smaller than the ion Debye length since $\alpha$ is of order
unity [see Fig.1(b)] and $\bar{Z}=Z_{d}T_{i}/T_{d}>>1.$

\subsection{Region 2 solution: $1>\psi>1/\bar{Z}$}

The $\exp\left(  -\bar{Z}\psi\right)  $ term is dropped from
Eq.(\ref{grand PE}) in region 2 because $\bar{Z}\psi$ is large. Taking into
account $T_{i}\ <<T_{e}$ and $\psi<1,$ Eq.(\ref{grand PE}) reduces to%
\begin{equation}
\frac{1}{\bar{r}^{2}}\frac{\partial}{\partial\bar{r}}\left(  \bar{r}^{2}%
\frac{\partial\psi}{\partial\bar{r}}\right)  =\ \underset{\text{ions}%
}{\underbrace{1+\psi}}-\underset{\text{electrons}}{\underbrace{\left(
1-\alpha\right)  }} \label{PE region2}%
\end{equation}
or
\begin{equation}
\frac{1}{\bar{r}^{2}}\frac{\partial}{\partial\bar{r}}\left(  \bar{r}^{2}%
\frac{\partial\psi}{\partial\bar{r}}\right)  =\psi+\alpha. \label{PE5}%
\end{equation}
By considering \ $\psi+\alpha$ as the unknown, it is seen that $\psi+\alpha$
has solutions of the form $\bar{r}^{-1}\exp(\pm\bar{r});$ the exponentially
growing solution is allowed here because region 2 does not extend to infinity.
A convenient way of expressing the general solution is
\begin{equation}
\psi+\alpha=\frac{A\cosh\left(  \bar{r}-\bar{r}_{o}\right)  +B\sinh\left(
\bar{r}-\bar{r}_{o}\right)  }{\bar{r}}. \label{psi2 gen}%
\end{equation}
In order to have $\psi=1/\bar{Z}$ when $\bar{r}=\bar{r}_{o}$, we choose%
\begin{equation}
\frac{1}{\bar{Z}}+\alpha=\frac{A}{\bar{r}_{o}} \label{A}%
\end{equation}
and leave $B$ undetermined. Thus%
\begin{equation}
\psi_{2}\ =\frac{\bar{r}_{o}\left(  \dfrac{1}{\bar{Z}}+\alpha\right)
\cosh\left(  \bar{r}-\bar{r}_{o}\right)  +B\sinh\left(  \bar{r}-\bar{r}%
_{o}\right)  -\alpha\bar{r}}{\bar{r}}\ \label{psi2}%
\end{equation}
$\allowbreak$ $\allowbreak$is the region 2 solution with coefficients arranged
so that $\psi_{2}=1/\bar{Z}$ when $\bar{r}=\bar{r}_{o}.$

\subsection{Region 1:\ Inner region\textbf{ }}

In this region $\psi>1$ and we assume that $\psi T_{i}/T_{e}<<1\,\ $so that
Eq.(\ref{grand PE}) reduces to
\begin{equation}
\frac{1}{\bar{r}^{2}}\frac{\partial}{\partial\bar{r}}\left(  \bar{r}^{2}%
\frac{\partial\psi}{\partial\bar{r}}\right)  =\ \underset{\text{ions}%
}{\underbrace{\frac{2}{\sqrt{\pi}}\sqrt{\psi}}}-\underset{\text{electrons}%
}{\underbrace{\left(  1-\alpha\right)  }}\ \simeq\ \underset{\text{ions}%
}{\underbrace{\frac{2}{\sqrt{\pi}}\sqrt{\psi}}}\ \label{region 1}%
\end{equation}
where the electron term has been dropped because $1-\alpha$ is significantly
less than unity and $\psi$ is assumed to be larger than unity. Equation
(\ref{region 1}) can be written as%
\begin{equation}
\ \ \ \frac{\partial^{2}\psi}{\partial\bar{r}^{2}}+\frac{2}{\bar{r}}%
\frac{\partial\psi}{\partial\bar{r}}=\mu(\psi)\psi\label{mu equn}%
\end{equation}
where%
\begin{equation}
\mu(\psi)=\frac{2}{\sqrt{\pi\psi}}<<1. \label{mu}%
\end{equation}
Since $\mu<<1$, the right hand side of Eq.(\ref{mu equn}) may be neglected
compared to either of the left hand terms in which case the approximate
solution to Eq.(\ref{mu equn}) is the vacuum-like solution
\begin{equation}
\psi=\frac{c+d\bar{r}}{\bar{r}} \label{psi-homo}%
\end{equation}
where $c$ and $d$ are constants to be determined. The coefficient $d$ provides
for the slight depression of the grain potential due to the shielding cloud.
The $d$ term is allowed because region 1 is of finite extent and so finite $d$
is not inconsistent with $\psi$ vanishing at infinity since infinity is not
located in region 1.

From Gauss' law, the radial electric field $E_{r}\ $ at the dust grain surface
is
\begin{equation}
4\pi\varepsilon_{0}r_{d}^{2}E_{r}=-Z_{d}e. \label{sfc bc}%
\end{equation}
Since $E_{r}=-\partial\phi/\partial r=\left(  \kappa T_{i}/e\lambda
_{Di}\right)  \partial\psi/\partial\bar{r}$, the boundary condition at the
grain surface $r_{d}$ can be expressed as%
\begin{equation}
\ \left(  \frac{\partial\psi}{\partial\bar{r}}\right)  _{\bar{r}_{d}}%
=-\frac{Z_{d}\ }{4\pi n_{i0}\lambda_{Di}^{3}}\frac{1}{\bar{r}_{d}^{2}}.
\label{Gauss}%
\end{equation}
This gives%
\begin{equation}
c=\ \frac{Z_{d}\ }{4\pi n_{i0}\lambda_{Di}^{3}}\ \ \label{solve c}%
\end{equation}
\bigskip and so, using Eqs.(\ref{a-norm}) and (\ref{define alpha}),%

\begin{equation}
c=\frac{\alpha\bar{a}^{3}}{3}. \label{solve c better}%
\end{equation}
By assumption $\psi=1$ at $\bar{r}_{i}$ and so, using Eq.(\ref{psi-homo}),%
\begin{equation}
\ \frac{c\ }{\bar{r}_{i}}+d=1 \label{cd}%
\end{equation}
in which case%
\begin{equation}
d=1-\frac{c\ }{\bar{r}_{i}}. \label{d}%
\end{equation}
Thus the region 1 potential is
\begin{equation}
\psi_{1}=\frac{\dfrac{\alpha\bar{a}^{3}}{3}+\left(  1-\dfrac{1\ }{\bar{r}_{i}%
}\dfrac{\alpha\bar{a}^{3}}{3}\right)  \bar{r}}{\bar{r}}; \label{L unity inner}%
\end{equation}
this satisfies Gauss's law at the dust grain surface and also gives $\psi=1$
at $\bar{r}=\bar{r}_{i}.$ The potential on the grain surface is%
\begin{equation}
\psi_{d}=\frac{\alpha\bar{a}^{3}}{3}\left(  \frac{1}{\bar{r}_{d}}-\dfrac
{1\ }{\bar{r}_{i}}\right)  +1.\ \label{psi-d}%
\end{equation}

\section{Matching the solutions}

\subsection{Matching principle}

Matching consists of arranging for equality of $\psi$ and $\psi^{\prime}$ at
the two interfaces between the three collisionless regions; the necessity for
continuity \ of $\psi$ and $\psi^{\prime}$ across an interface is established
by integrating Eq.(\ref{grand PE}) twice across the interface. Solutions on
the left and right hand sides of a matching radius $\bar{r}_{m}$ are of the
general form $\psi_{left}=L(\bar{r})/\bar{r}$ and $\psi_{right}=R(\bar
{r})/\bar{r}$ $\,$and so matching requires%
\begin{align}
L(\bar{r}_{m})/\bar{r}_{m}  &  =\psi_{m}=R(\bar{r}_{m})/\bar{r}_{m}%
\label{match function}\\
L^{\prime}(\bar{r}_{m})  &  =R^{\prime}(\bar{r}_{m}) \label{match derivative}%
\end{align}
where $\psi_{m}=1/\bar{Z}$ when $\bar{r}_{m}=\bar{r}_{o}$ and $\psi_{m}=1$
when $\bar{r}_{m}=\bar{r}_{i}$. Here $L$ and $R$ are the left and right hand
numerators: $L$ is the numerator of $\psi_{1}$ and $R$ is the numerator of
$\psi_{2}$ when $\bar{r}_{m}=\bar{r}_{i};$ $L$ is the numerator of $\psi_{2}$
and $R$ is the numerator of $\psi_{3}$ when $\bar{r}_{m}=\bar{r}_{o}.$

\subsection{Matching of $\psi_{2}$ and $\psi_{3}$ at $\bar{r}_{o}$}

The $\psi_{2}$ and $\psi_{3}$ solutions have already been arranged to satisfy
Eq.(\ref{match function}) (i.e., $\psi_{2}=\psi_{3}=1/\bar{Z}$ at $\bar{r}%
_{o}$). The derivative matching condition, Eq.(\ref{match derivative}), is
satisfied if
\begin{equation}
B=\alpha-\frac{r_{o}}{\bar{Z}}\sqrt{\alpha\bar{Z}\ +1\ } \label{solve B}%
\end{equation}
and so
\begin{equation}
\psi_{2}\ =\frac{\bar{r}_{o}\left(  \dfrac{1}{\bar{Z}}+\alpha\right)
\cosh\left(  \bar{r}-\bar{r}_{o}\right)  +\left(  \alpha-\dfrac{\bar{r}_{o}%
}{\bar{Z}}\sqrt{\alpha\bar{Z}\ +1\ }\right)  \sinh\left(  \bar{r}-\bar{r}%
_{o}\right)  -\alpha\bar{r}}{\bar{r}} \label{psi2 solution}%
\end{equation}
smoothly matches to $\psi_{3}$ at $\bar{r}_{o}.$ The actual value of $\bar
{r}_{o}$ is undetermined at this stage and will be found later.

\bigskip

\subsection{Matching of $\psi_{1}$ and $\psi_{2}$ at $\bar{r}_{i}$}

Since Eq.(\ref{match function}) requires $R=\bar{r}_{i}$ in order to have
$\psi(\bar{r}_{i})=1$, Eq.(\ref{psi2 solution}) provides the relation%
\begin{equation}
\bar{r}_{i}=\bar{r}_{o}\left(  \dfrac{1}{\bar{Z}}+\alpha\right)  \cosh\left(
\bar{r}_{i}-\bar{r}_{o}\right)  +\left(  \alpha-\dfrac{\bar{r}_{o}}{\bar{Z}%
}\sqrt{\alpha\bar{Z}\ +1\ }\right)  \sinh\left(  \bar{r}_{i}-\bar{r}%
_{o}\right)  -\alpha\bar{r}_{i}. \label{L equal R}%
\end{equation}
The condition $L=\bar{r}_{i}$ when $\psi=1$ has already been arranged by the
form of Eq.(\ref{L unity inner}).

The condition $L^{\prime}=R^{\prime}$, found by taking derivatives of the
numerators of Eqs.(\ref{L unity inner}) and (\ref{psi2 solution}), is%

\begin{equation}
1-\ \dfrac{\alpha\bar{a}^{3}}{3\bar{r}_{i}}=\bar{r}_{o}\left(  \dfrac{1}%
{\bar{Z}}+\alpha\right)  \sinh\left(  \bar{r}_{i}-\bar{r}_{o}\right)  +\left(
\alpha-\dfrac{\bar{r}_{o}}{\bar{Z}}\sqrt{\alpha\bar{Z}\ +1\ }\right)
\cosh\left(  \bar{r}_{i}-\bar{r}_{o}\right)  -\alpha.
\label{Lprime equal Rprime}%
\end{equation}

Equations (\ref{L equal R}) and (\ref{Lprime equal Rprime}) constitute two
coupled equations in the unknowns $\bar{r}_{i}$ and $\bar{r}_{o}.$ Using
$\bar{Z}>>1,$ these equations reduce to$\ $
\begin{align}
\ \left(  1+\alpha\right)  \bar{r}_{i}  &  =\alpha\bar{r}_{o}\cosh\left(
\bar{r}_{i}-\bar{r}_{o}\right)  +\alpha\sinh\left(  \bar{r}_{i}-\bar{r}%
_{o}\right) \label{coupled-1}\\
1+\alpha-\dfrac{\alpha\bar{a}^{3}\ }{3\bar{r}_{i}}  &  =\alpha\bar{r}_{o}%
\sinh\left(  \bar{r}_{i}-\bar{r}_{o}\right)  +\alpha\cosh\left(  \bar{r}%
_{i}-\bar{r}_{o}\right)  . \label{coupled-2}%
\end{align}
For given $\bar{a}$ and $\alpha$ these nonlinear equations can be solved
numerically for $\bar{r}_{i}$ and $\bar{r}_{o}.$ Since $\alpha=\alpha(\bar
{a},\bar{r}_{d}),$ this means that for any point in $\bar{a},\bar{r}_{d}$
parameter space, one can calculate $\alpha$ and then calculate $\bar{r}_{i}$
and $\bar{r}_{o}$. Thus, we can consider $\bar{r}_{i}=\bar{r}_{i}(\bar{a}%
,\bar{r}_{d})$ and $\bar{r}_{o}=\bar{r}_{o}(\bar{a},\bar{r}_{d}).$

Once $\bar{r}_{i}$ and $\bar{r}_{o}$ are known, the solutions $\psi_{1}%
,\psi_{2},$ and $\psi_{3}$ are all determined and match smoothly across the
interfaces. The Vlasov/Poisson equation is thus solved all the way from the
grain surface to infinity. The potential falls off abruptly at $\bar{r}%
>\bar{r}_{o}$ with a scale length given by the dust Debye length. The dust
Debye length is thus of physical importance even though it is much smaller
than the inter-particle spacing. No paradoxes occur due to this situation
because the solution for $\psi$ is multi-scale and more complicated than a
simple Yukawa type potential. In particular, the dust shielding does not take
place in a sphere having a radius equal to the dust Debye length, but instead
takes place over the surface of a sphere having a much larger radius (a few
times the ion Debye length). The extremely sharp cut-off of $\psi$ \ beyond
$\bar{r}_{o}$ completely decouples dust grains from each other if their
interparticle separation distance exceeds $\bar{r}_{o}.$%

%TCIMACRO{\FRAME{fhFU}{4.7986in}{5.3873in}{0pt}{\Qcb{Comparison of experiment
%parameters (short solid line marked `expt') with model prediction that dusty
%plasma will crystallize if experiment parameters intersect the condensation
%curve or are above it. }}{}{new-049406php3.eps}%
%{\special{ language "Scientific Word";  type "GRAPHIC";
%maintain-aspect-ratio TRUE;  display "USEDEF";  valid_file "F";
%width 4.7986in;  height 5.3873in;  depth 0pt;  original-width 6.7173in;
%original-height 7.5475in;  cropleft "0";  croptop "1";  cropright "1";
%cropbottom "0";  filename 'NEW-049406PHP3.eps';file-properties "XNPEU";}}}%
%BeginExpansion
\begin{figure}
[h]
\begin{center}
\includegraphics[
height=5.3873in,
width=4.7986in
]%
{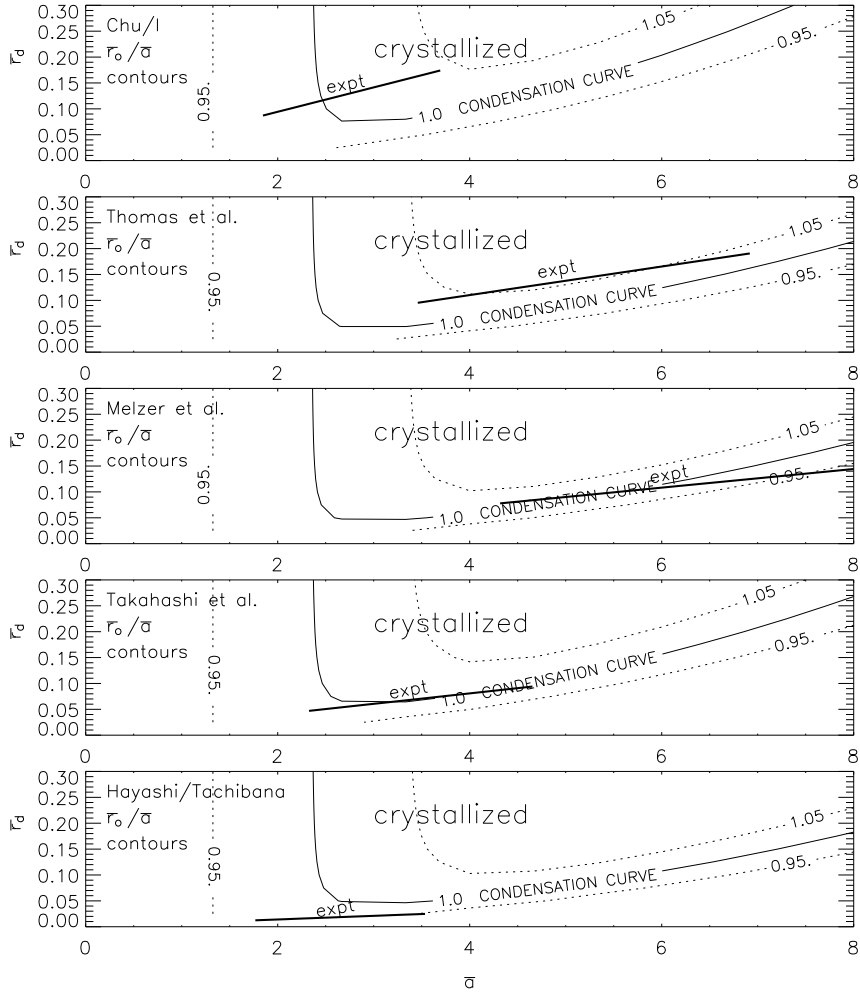}%
\caption{Comparison of experiment parameters (short solid line marked `expt')
with model prediction that dusty plasma will crystallize if experiment
parameters intersect the condensation curve or are above it. }%
\end{center}
\end{figure}
%EndExpansion

\section{\textbf{Crystallization }}

When $\bar{a}>\bar{r}_{o}$ each dust grain is decoupled from neighboring dust
grains and so the dust grains behave as a gas of non-interacting particles.
However, if $\bar{a}<\bar{r}_{o},$ then each dust grain is within the
shielding cloud of its neighbor and subject to the unshielded repulsive force
of its neighbor. Because the repulsive force scales as $\bar{Z}\psi,$ this
repulsion becomes enormous as soon as a dust grain tries to move any
significant distance inside of the $\bar{r}=\bar{r}_{o}$ layer (i.e., inside
of the $\psi=1/\bar{Z}$ layer). When experienced by other dust grains, a test
particle dust grain thus acts like a hard sphere with radius $\bar{r}_{o}.$
Thus, dust grains cannot move significantly inside $\bar{r}_{o}$ and so the
condition for strong coupling and crystallization is that $\bar{a}$ becomes
less than $\bar{r}_{o}.$ The condensation curve is found by making the
following sequence of calculations at each point $\bar{a},\bar{r}_{d}$ in
dusty plasma parameter space (i) calculate $\alpha(\bar{a},\bar{r}_{d}),$ (ii)
calculate $\bar{r}_{i}$ and $\bar{r}_{o}$ by solving the nonlinear coupled
Eqs.(\ref{coupled-1}) and (\ref{coupled-2}), $\ $(iii) plot the locus of the
curve $\bar{r}_{o}=\bar{a}$ and establish which side of this curve corresponds
to $\bar{a}<\bar{r}_{o}.$ The uppermost plot in Fig. 3 shows contours of
constant $\bar{r}_{o}/\bar{a}$ calculated for Ref.\cite{Chu:1994} and marks
the contour where $\bar{r}_{o}/\bar{a}=1$ as the `condensation curve'; above
the condensation curve (and with $\bar{r}_{d}<<1$ as discussed earlier) the
dusty plasma should be crystallized. It is seen that portions of the
experiment line lie above the condensation curve which means that the model
predicts that the dusty plasma of Ref. \cite{Chu:1994} \ should be
crystallized. Thus, there is excellent agreement between the model and the
experimental parameters of Ref. \cite{Chu:1994}.

The other plots in Fig.3 are similar, but use data from the experimental
results reported by Thomas et al. \cite{Thomas:1994}, Melzer et al.
\cite{Melzer:1994}, Hayashi and Tachibana \cite{Hayashi:1994}, and Takahashi
et al. \cite{Takahashi:1998}. There is excellent agreement between the model
and all these experiments with the exception of the Hayashi/Tachibana
experiment where the experimental curve lies slightly below the condensation
curve. The upper part of Table 1 lists the parameters of these experiments
while the lower part gives the results of dust charging theory and then the
results of this model using a best-fit density that is within experimental
error. The main result is the values of $\bar{r}_{i}$ and $\bar{r}_{o}$. The
$\bar{r}_{i}$ \ values in Table I greatly exceed $\bar{r}_{d}$ showing that
the depression of the grain surface potential due to shielding is only a
slight effect. Table 1 shows that when the published parameters of the
experiments are used, a value of $\bar{r}_{o}$ is calculated which is slightly
larger than $\bar{a}$; the calculated ratio $\bar{r}_{o}/\bar{a}$ is given in
the bottom line of Table 1. The fact that $\bar{r}_{o}/\bar{a}$ is greater
than unity indicates that the experiment is above the condensation curve and
so should be crystallized --- this is our main result (the slight disagreement
of the Hayashi/Tachibana experiment will be discussed later). \ For reference,
Table 1 also lists the value of $\ \Gamma$ associated with these experiments,
and it is interesting to note that according to our model $\Gamma$ has no
physical significance regarding condensation and so it is not surprising that
$\Gamma$ has a range of quite different values for the different experiments.

It has not been possible to compare the model to the experiment underway
\cite{PKE:2003} on board the International Space Station, because plasma
densities and temperatures have not yet been provided for that experiment.

Figure 4 shows plots of $\ $ log$\psi,$ $\psi$, $\psi$ on an expanded scale
(to show the behavior when $\psi\sim1/\bar{Z}$)$,$ $n_{e}/n_{e0},$
$n_{i}/n_{i0},$ and $n_{d}/n_{d0}$ for the Chu and I\ experiment
\cite{Chu:1994} using the values of $\bar{r}_{i}$ and $\bar{r}_{o}$ listed in
Table 1. The $\psi(\bar{r})$ plotted in Fig. 4 is calculated using
Eq.(\ref{L unity inner}) in region 1, Eq.(\ref{psi2 solution}) in region 2,
and Eq.(\ref{small psi soln}) in region 3; the electron, ion and dust
densities in Fig. 4 are calculated using Eqs.(\ref{ne}), (\ref{ni}), and
(\ref{nd}) respectively. The dust temperature $T_{d}$ has been assumed to
equal $T_{i}$ so that $\bar{Z}=Z_{d};$ different values of $T_{d}$ would only
change the decay rate of $\psi$ outside of $\bar{r}_{o}\,,$ but would not
change the values of $\bar{r}_{i}$ and $\bar{r}_{o}\ $since these are
insensitive to the value of $\bar{Z}$ as long as it is large compared to
unity. It is seen that there is a sharp cut-off of the potential at $\bar
{r}_{o}\ $ and that, beyond this radius, the potential decays precipitously
with a characteristic scale length given by the dust Debye length. The
potential curve is smooth all the way from the dust grain surface to infinity;
this smoothness results from choosing $\bar{r}_{i}$ and $\bar{r}_{o}$ to match
$\psi$ and its derivatives at the interfaces between collisionless regions.%

%TCIMACRO{\FRAME{fhFU}{2.6027in}{5.5757in}{0pt}{\Qcb{Solutions for nominal
%parameters of Chu and I experiment as a function of $\bar{r}.$ As shown in
%Table 1, the relevant parameters are $T_{e}=$ $2$ eV, $T_{i}=0.03$ eV,
%$\bar{a}=2.61,$ $\bar{r}_{d}=0.12$, $\bar{r}_{i}=1.51,$ $\bar{r}_{o}\ =2.63,$
%$\alpha=0.96,$ and $\bar{Z}=4.8\times10^{3}.$ From top to bottom plots are:
%$\log_{10}\psi,$ $\psi,\ 10^{4}\psi$ which gives an expanded scale to show the
%region 3 decay, $n_{e}/n_{e0},$ $n_{i}/n_{i0},$ and $n_{d}/n_{d0}.$ The
%functional form of $\psi$ is determined from the appropriate asymptotic form
%in each of the three regions. Note the sharp cut-off of the dust density at
%$\bar{r}=\bar{r}_{o};$ the scale length of this cutoff is the dust Debye
%length $\lambda_{Dd}.$}}{}{049406php4.eps}%
%{\special{ language "Scientific Word";  type "GRAPHIC";
%maintain-aspect-ratio TRUE;  display "USEDEF";  valid_file "F";
%width 2.6027in;  height 5.5757in;  depth 0pt;  original-width 3.3433in;
%original-height 7.2046in;  cropleft "0";  croptop "1";  cropright "1";
%cropbottom "0";  filename '049406PHP4.eps';file-properties "XNPEU";}}}%
%BeginExpansion
\begin{figure}
[h]
\begin{center}
\includegraphics[
height=5.5757in,
width=2.6027in
]%
{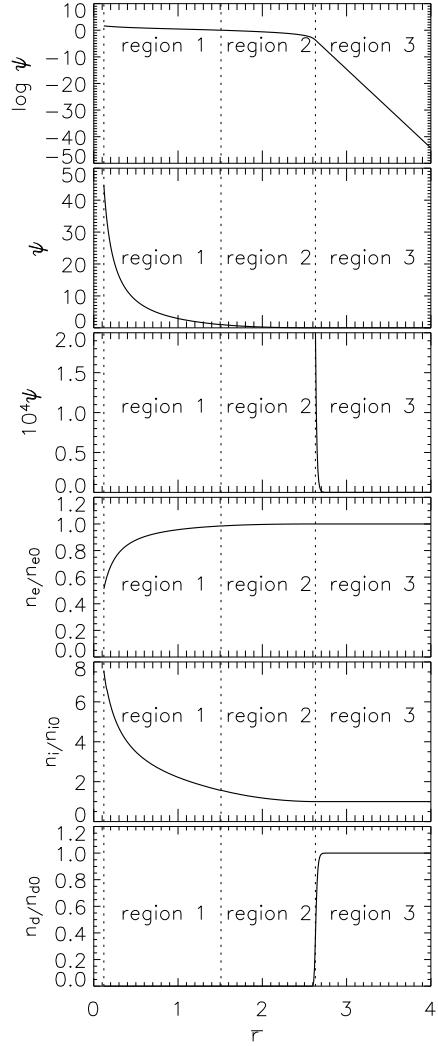}%
\caption{Solutions for nominal parameters of Chu and I experiment as a
function of $\bar{r}.$ As shown in Table 1, the relevant parameters are
$T_{e}=$ $2$ eV, $T_{i}=0.03$ eV, $\bar{a}=2.61,$ $\bar{r}_{d}=0.12$, $\bar
{r}_{i}=1.51,$ $\bar{r}_{o}\ =2.63,$ $\alpha=0.96,$ and $\bar{Z}%
=4.8\times10^{3}.$ From top to bottom plots are: $\log_{10}\psi,$
$\psi,\ 10^{4}\psi$ which gives an expanded scale to show the region 3 decay,
$n_{e}/n_{e0},$ $n_{i}/n_{i0},$ and $n_{d}/n_{d0}.$ The functional form of
$\psi$ is determined from the appropriate asymptotic form in each of the three
regions. Note the sharp cut-off of the dust density at $\bar{r}=\bar{r}_{o};$
the scale length of this cutoff is the dust Debye length $\lambda_{Dd}.$}%
\end{center}
\end{figure}
%EndExpansion

\bigskip

A question arises regarding why the Hayashi/Tachibana experiment
\cite{Hayashi:1994} lies slightly below the condensation curve (i.e., has
$\bar{r}_{o}/\bar{a}=0.97$ rather than above unity). Examination of the
parametric sensitivity of the model predictions shows that intersection of the
Hayashi/Tachibana experiment with the predicted condensation curve occurs if
the assumed electron temperature is increased to $T_{e}=8$ eV or if the
assumed grain diameter is doubled. Increasing the assumed atomic mass number
to values larger than 16 also causes the experiment curve to approach the
condensation curve but this effect is minimal in the relevant parameter range.
Reference \cite{Hayashi:1994} reported an ion temperature which was not
measured, but assumed, and an electron temperature which was estimated based
on earlier measurements \cite{Tachibana:1984} made in another plasma under
similar conditions. It is possible therefore that the slight discrepancy
between the model predictions and the Hayashi/Tachibana experiment results
from an inaccurate estimation of the electron to ion temperature ratio in the
region of the dust grains. Better agreement would be obtained with a higher
$T_{e}/T_{i}$ ratio and the values of $T_{e}$ and $T_{i}$ used in Table I were
chosen to correspond to room temperature ions and the electron temperature
measurement given in Fig.4 of Ref.\cite{Tachibana:1984}. Comparison with a
scanning electron microscope measurement has shown that the Mie scattering
technique used by Hayashi/Tachibana to measure the dust grain diameter is
quite accurate \cite{Tachibana:1995}, so it is unlikely that the discrepancy
between the model predictions and the Hayashi/Tachibana experiment is due to a
factor of two error in measurement of the dust grain diameter.

Another question to be addressed is the\ possible importance of (i)\ barriers
due to local maxima in the effective potential
\cite{Bernstein:1959,Lampe:2000} and (ii) ion trapping/detrapping due to
collisions\cite{Goree:1992,Zobnin:2000,Lampe:2003}. The model presented here
argues that the amount of net charge in region 1 is so small that the
potential in region 1 is nearly the same as the vacuum potential that would be
produced by a bare, unshielded dust grain [see Eq.(\ref{psi-homo})]. Changing
the amount of charge in region 1 by factors as large as order unity would not
affect this argument. Thus any reduction in the amount of charge in region 1
because of effective potential barriers will make no difference to the region
1 solution, because it is already assumed that there is no charge in region 1.
So long as the number of trapped ions in region 1 is small compared to
$Z_{d},$ the potential in region 1 is mainly due to the dust charge and again
it is reasonable to use the vacuum potential in region 1. As for region 2, the
ion density predicted by collisionless theory in region 2 is identical to the
linearized Boltzmann relation obtained from collisional theory [see
Eq.(\ref{PE region2})]. Since trapping and detrapping result from collisions,
trapping and detrapping should tend to make the system more Boltzmann-like,
but since the system is already Boltzmann-like in region 2, trapping and
detrapping should not cause significant changes to the region 2 ion density
profile and thus should not significantly affect the $\psi$ profile in region
2. Effective potential barriers in region 2 may rearrange the radial charge
distribution in region 2 slightly, but this should cause only a small effect
on $\psi$ because $\psi$ is a double integral with respect to radius of the
net charge distribution [see Eq.(\ref{Poisson norm})]. As for region 3, the
normalized potential $\psi$ is so small in region 3 $\ $that ions are
unaffected by any spatial dependence of the potential; any corrections to the
region 3 potential profile should therefore have negligible effect on ion
trajectories. Thus, while effective potential barriers and trapping/detrapping
may modify the net charge radial profile somewhat, these should have a much
reduced effect on the $\psi$ profile and so should not cause any substantial
changes in the values of $\bar{r}_{i}$ or $\bar{r}_{o}.$ Small changes in the
$\psi$ profile should not affect the basic premise that there exist three
concentric collisionless regions each with distinct physics nor \ the
conclusion that dust grains condense when the radius $\bar{r}_{o}$ of the
interface between regions 2 and 3 exceeds $\bar{a},$ the nominal intergrain
spacing distance.

At this point in the discussion it is possible to revisit the assumption made
at the end of Section III that ion capture by dust grains may be ignored when
characterizing the collisionless ion velocity distribution function in regions
1-3. Ignoring ion capture by the dust grain is tantamount to saying that all
ions entering the collisionless region are reflected radially so that there
are equal numbers of ions moving radially inward and outwards; if some ions
were captured by the dust grain, there would be fewer ions moving radially
outwards than inwards. The number of captured ions can be estimated using
OML\ theory \cite{Shukla:ionXsection} which shows that the effective
cross-section for ions entering from a radius where the potential is zero and
then being captured by a dust grain is $\sigma_{capture}\sim(1+\psi_{d}%
)\pi\bar{r}_{d}^{2}$. This capture cross-section is to be compared to
$\sigma_{enter}=\pi\bar{r}_{o}^{2}$ the cross-section for ions to enter region
2 from outside (the outer boundary of region 2 is used because this denotes
the edge of the potential well seen by the ions). Of\ the ions that enter
region 2, the fraction that are captured by the dust grain is given by the
ratio $\sigma_{capture}/\sigma_{enter}\simeq$ $(1+\psi_{d})\bar{r}_{d}%
^{2}/\bar{r}_{o}^{2}$. Evaluation of this ratio using the Chu and I parameters
in Table I ($\psi_{d}=46,$ $\bar{r}_{d}=0.12,$ $\bar{r}_{0}=2.63$) gives
$\sigma_{capture}/\sigma_{enter}\simeq0.1$ which shows that the fraction of
ions captured is small enough to be neglected (similar results hold for the
other experiments listed in Table I). This validates the assumption made in
Section III that distinctions\ \cite{Allen:1956,Bernstein:1959} between the
outward and inward ion velocity distributions \thinspace$f_{+},$ $f_{-}$ may
be neglected and confirms that Eq.(\ref{f soln}) is a suitable representation
for the ion distribution function.

\section{Summary}

The standard linear fluid analysis of Debye shielding fails\ when $\left\vert
q\phi/\kappa T\right\vert $ exceeds unity because the linear Debye shielding
model is based on the assumption that $\left\vert q\phi/\kappa T\right\vert $
is small compared to unity. This issue is important for condensation of dusty
plasmas, because condensation requires having $\left\vert Z_{d}e\phi/\kappa
T_{d}\right\vert $ exceed unity.

Dusty plasmas can be characterized by an $\bar{a},$ $\bar{r}_{d}$ parameter
space where $\bar{a}$ and $\bar{r}_{d}$ are the inter-grain spacing distance
and grain radius normalized to the ion Debye length. An experiment corresponds
to a point in this parameter space and if the density of the experiment is not
known precisely, then the range of densities within experimental error
corresponds to a slanted line segment in this parameter space.

Because shielding distances are much smaller than an ion collision mean free
path, ions can be considered as collisionless in the shielding sphere
surrounding a dust grain. A collisionless Vlasov model is used to calculate
particle densities in the electrostatic potential of a dust grain test charge.
This collisionless theory gives the same results as does Boltzmann theory for
electrons and for dust grains because they are negatively charged but gives
results different from Boltzmann theory for ions in the vicinity of the dust
grain. Ions near a highly charged dust grain test particle fall into a deep
potential well and are accelerated to high velocities. This means that no ions
have zero velocity near the dust grain test particle and so integrals over the
velocity distribution have a lower velocity limit corresponding to the minimum
velocity of an ion falling into the deep potential well. This invalidates the
fluid theory concept of pressure because pressure is based on the assumption
of the existence of random velocities about some mean. For $\left\vert
e\phi/\kappa T_{i}\right\vert <<1,$ the ion density corresponds to the
Boltzmann prediction, but for $\left\vert e\phi/\kappa T_{i}\right\vert >>1$
the ion density is much less than that predicted by the Boltzmann relation.

The potential in the vicinity of a dust grain test particle has three distinct
types of behavior. These behaviors occur in concentric spherical regions
consisting of (1)\ \ an inner region where the potential is vacuum-like,
(2)\ a middle region where the potential includes both growing and decaying
Yukawa-like terms with characteristic scale lengths of the order of the ion
Debye length, and (3)\ an outer region with a rapidly decaying Yukawa-type
solution having a scale length of the order of the dust Debye length. Region 1
physics differs from fluid theory, is consistent with dust charging physics,
and avoids the paradoxes intrinsic to fluid theory at large $\left\vert
e\phi/\kappa T\right\vert $.

For any point $\bar{a},$ $\bar{r}_{d}$ in dusty plasma parameter space, the
requirement for smooth matching of the solutions at the interfaces between the
three inner regions determines the locations $\bar{r}_{i}$ and $\bar{r}%
_{o}\,\ $of these interfaces. Condensation occurs when $\bar{a}<\bar{r}_{o}$
and occurs on the line $\bar{a}=\bar{r}_{o}\left(  \bar{a},\bar{r}_{d}\right)
$ which gives a `condensation curve' in dusty plasma parameter space; $\bar
{a}$ is less than $\bar{r}_{o}$ above this curve and in this region the dusty
plasma is crystallized. The model predicts condensation parameters in good
agreement with published experiments.

\pagebreak

\pagebreak%

%TCIMACRO{\TeXButton{B}{\begin{table}[tbp] \centering}}%
%BeginExpansion
\begin{table}[tbp] \centering
%EndExpansion
\caption{
Comparison between model predictions and experiments{}}%
\begin{tabular}
[c]{lllllll}\hline\hline
First author \& reference &  & Chu \ \cite{Chu:1994} & Thomas
\cite{Thomas:1994} & Melzer \cite{Melzer:1994} & Takahashi
\cite{Takahashi:1998} & Hayashi \cite{Hayashi:1994}\\\hline
reported value & $n_{i0}($cm$^{-3})$ & $10^{9}$ & $10^{9}$ & $2\times10^{8}$ &
$10^{9}$ & $10^{9}$\\
" & $n_{d0}($cm$^{-3})$ & $2\times10^{5}$ & $4\times10^{4}$ & $1.4\times
10^{3}$ & $10^{5}$ & $3\times10^{5}$\\
" & $r_{d}$ ($\mu$m) & $5$ & $5$ & $12.5$ & $5.4$ & $1.3$\\
" & $T_{e}($eV$)$ & $2$ & $3$ & $4$ & $3$ & $4.4$\\
" & $T_{i}($eV$)$ & $0.03$ & $0.025$ & $0.03$ & $0.03$ & $0.025$\\
from Eq.(\ref{Wigner-Seitz}) & $\,a$ ($\mu$m) & $106$ & $181$ & $554$ & $133$
& $93$\\
neutral pressure & $\ $Pa & $16$ & $200$ & $80$ & $87$ & $0.3$\\
ion mean free path & $l_{mfp}$ ($\mu$m) & $5\times10^{3}$ & $6\times10^{2}$ &
$1.6\times10^{3}$ & $1.5\times10^{3}$ & $3\times10^{3}$\\
ion mass & amu & $40$ & $40$ & $16$ & $26$ & $16$\\
&  &  &  &  &  & \\
\textbf{Modeled quantities:} &  &  &  &  &  & \\
ion density used in model & $n_{i0}($cm$^{-3})$ & $10^{9}$ & $10^{9}$ &
$10^{8}$ & $10^{9}$ & $10^{9}$\\
from Eq.(\ref{Debye}) & $\lambda_{di}$ ($\mu$m) & $41$ & $37$ & $128$ & $41$ &
$37$\\
& $4\pi n_{i0}\lambda_{Di}^{3}$ & $842$ & $641$ & $2663$ & $842$ & $641$\\
from Eq.(\ref{a-norm}) & $\bar{a}\ $ & $2.61$ & $4.89$ & $4.31$ & $3.29$ &
$2.50$\\
& $\bar{r}_{d}$ & $0.12$ & $0.13$ & $0.078$ & $0.066$ & $0.018$\\
from Eq.(\ref{charging param}) & $P$ & $0.021$ & $0.0035$ & $0.0029$ &
$0.0056$ & $0.0034$\\
from Eq.(\ref{P dep}) & $\psi_{d}$ & $46$ & $194$ & $219$ & $135$ & $208$\\
from Eq.(\ref{define alpha}) & $\alpha$ & $0.96$ & $0.67$ & $0.64$ & $0.76$ &
$0.70$\\
from Eq.(\ref{Zd}) & $Z_{d}$ & $4.8\times10^{3}$ & $1.6\times10^{4}$ &
$4.5\times10^{4}$ & $7.6\times10^{3}$ & $2.3\times10^{3}$\\
$\Gamma$ & $\dfrac{Z_{d}^{2}e^{2}}{4\pi\varepsilon_{0}a\kappa T_{i}}$ &
$10^{4}$ & $8.9\times10^{4}$ & $1.8\times10^{5}$ & $2.1\times10^{4}$ &
$3.4\times10^{3}$\\
solution of Eqs.(\ref{coupled-1}),(\ref{coupled-2}) & $\bar{r}_{i}$ & $1.51$ &
$3.76$ & $2.99$ & $2.03$ & $1.20$\\
solution of Eqs.(\ref{coupled-1}),(\ref{coupled-2}) & $\bar{r}_{o}$ & $2.63$ &
$5.18$ & $4.41$ & $3.31$ & $2.42$\\
& $\bar{r}_{o}/\bar{a}$ & $1.005$ & $1.06$ & $1.02$ & $1.005$ & $0.97$%
\\\hline\hline
\end{tabular}
\noindent%
%TCIMACRO{\TeXButton{E}{\end{table}}}%
%BeginExpansion
\end{table}%
%EndExpansion


\begin{thebibliography}{99}                                                                                               %


\bibitem {Shukla:2002}P. K. Shukla and A. A. Mamun, \textit{Introduction to
Dusty Plasma Physics}, (Institute of Physics Publishing Bristol and
Philadelphia, 2002).

\bibitem {Ikezi:1986}H. Ikezi, Phys. Fluids \textbf{29}, 1764 (1986).

\bibitem {Chu:1994}J. H. Chu and L. I, Phys. Rev. Lett. \textbf{72}, 4009 (1994).

\bibitem {Thomas:1994}H. Thomas, G. E. Morfill, V. Demmel, J. Goree, B.
Feuerbacher, and D. Mohlmann, Phys. Rev. Lett. \textbf{73}, 652 (1994).

\bibitem {Hayashi:1994}H. Hayashi and K. Tachibana, Japan J. Appl. Phys.
\textbf{33}, 804 (1994).

\bibitem {Melzer:1994}A. Melzer, T. Trottenberg, and A. Piel, Phys. Lett. A
\textbf{191, }301 (1994).

\bibitem {Morfill:2002}G. E. Morfill, B. M. Annaratone, P. Bryant, A. V.
Ivlev, H. M. Thomas, M. Zuzic, and V. E. Fortov, Plasma Phys. Control. Fusion
\textbf{44}, B263 (2002).

\bibitem {PKE:2003}A. P. Nefedov, G. E. Morfill, V. E. Fortov, H. M. Thomas,
H. Rothermel, T. Hagl, A. V. Ivlev, M. Zuzic, B. A. Klumov, A. M. Lipaev, V.
I. Molotkov, O. F. Petrov, Y. P. Gidzenko, S. K. Krikalev, W. Shepherd, A. I.
Ivanov, M. Roth, H. Binnenbruck, J. A. Goree and Y. P. Semenov, New J. Phys.
\textbf{5}, art. 33 (2003)

\bibitem {Slattery:1980}W. L. Slattery, G. D. Doolen, and H. E. Dewitt, Phys.
Rev. E \textbf{21}, 2087 (1980).

\bibitem {Wigner:1938}E. Wigner, Trans. Faraday Soc. \textbf{34,}768 (1938).

\bibitem {Nicholson}D. R. Nicholson, \textit{Introduction to Plasma Theory}
(Krieger Publishing Company, Malabar, Florida, 1992), Section 9.2.

\bibitem {Lampe:2000}M. Lampe, G. Joyce, and G. Ganguli, Phys. Plasmas
\textbf{7},2851 (2000).

\bibitem {Hansen:1995}C. Hansen and J. Fajans, Phys. Rev. Lett. \textbf{74},
4209 (1995).

\bibitem {Goree:1992}J. Goree, Phys. Rev. Letters \textbf{69}, 277 (1992).

\bibitem {Zobnin:2000}A. V. Zobnin, A. P. Nefedov, V. A. Sinel'shchikov, and
V. E. Fortov, JETP \textbf{91}, 483 (2000).

\bibitem {Lampe:2003}M. Lampe, R. Goswami, Z. Sternovsky, S. Robertson, V.
Gavrishchaka, G. Ganguli, and G. Joyce, Phys. Plasmas \textbf{10}, 1500 (2003).

\bibitem {Wang:1995}X. Wang and A. Bhattacharjee, Phys. Plasmas \textbf{3},
1189 (1996).

\bibitem {Otani:1997}N. Otani and A. Bhattacharjee, Phys. Rev. Lett.
\textbf{78}, 1468 (1997).

\bibitem {Mott:1926}H. M. Mott-Smith and I. Langmuir, Phys. Rev. \textbf{28,}
727 (1926).

\bibitem {Allen:1956}J. E. Allen, R. L. F. Boyd, and P. Reynolds, Proc. Phys.
Soc. \textbf{70}, 297 (1956).

\bibitem {Shukla:2002OML}P. K. Shukla and A. A. Mamun, \textit{ibid.},
Eq.(2.2.1), p. 38.

\bibitem {Shukla:2002a}P. K. Shukla and A. A. Mamun, \textit{ibid.},
Eq.(2.3.4), p. 51.

\bibitem {Watson:1956}K. M. Watson, Phys. Rev. \textbf{102}, 12 (1956).

\bibitem {Nicholson:motionconstants}D. R. Nicholson, \textit{Introduction to
Plasma Theory} (Krieger Publishing Company, Malabar, Florida, 1992), p.72.

\bibitem {Laframboise:1973}J. G. Laframboise and L. W. Parker, Phys. Fluids
\textbf{16}, 629 (1973).

\bibitem {Bernstein:1959}I. B. Bernstein and I. N. Rabinowitz, Phys. Fluids
\textbf{2}, 112 (1959).

\bibitem {Lampe:2001}M. Lampe, J. Plasma Phys. \textbf{65}, 171 (2001).

\bibitem {Takahashi:1998}K. Takahashi, T. Oishi, K. Shimomai, Y. Hayashi, and
S. Nishino, Phys. Rev. E \textbf{58}, 7805 (1998).

\bibitem {Tachibana:1984}K. Tachibana, M. Nishida, H. Harima, and Y. Urano, J.
Phys. D:\ Appl. Phys. \textbf{17}, 1727 (1984).

\bibitem {Tachibana:1995}K. Tachibana and Y. Hayashi, Aust. J. Phys.
\textbf{48}, 469 (1995).

\bibitem {Shukla:ionXsection}P. K. Shukla and A. A. Mamun, \textit{ibid.},
Eqs.(2.3.2) and (2.2.6) p. 38,39.
\end{thebibliography}
\end{document}